\documentclass[letterpaper,twocolumn,prl,superscriptaddress,showpacs]{revtex4}
\usepackage{graphics,graphicx,amsmath}
\usepackage{dcolumn}

\begin{document}
\newcommand{\zl}[2]{$#1\:\text{#2}$}
\newcommand{\zlm}[2]{$#1\:\mu\text{#2}$}
\title{Stopping Supersonic Beams with an Atomic Coilgun}
\date{\today}
\author{Edvardas Narevicius}
\affiliation{Center for Nonlinear Dynamics and Department of Physics, The University of Texas at Austin, Austin, Texas 78712-1081, USA}
\author{Adam Libson}
\affiliation{Center for Nonlinear Dynamics and Department of Physics, The University of Texas at Austin, Austin, Texas 78712-1081, USA}
\author{Christian G. Parthey}
\affiliation{Center for Nonlinear Dynamics and Department of Physics, The University of Texas at Austin, Austin, Texas 78712-1081, USA}
\author{Isaac Chavez}
\affiliation{Center for Nonlinear Dynamics and Department of Physics, The University of Texas at Austin, Austin, Texas 78712-1081, USA}
\author{Julia Narevicius}
\affiliation{Center for Nonlinear Dynamics and Department of Physics, The University of Texas at Austin, Austin, Texas 78712-1081, USA}
\author{Uzi Even}
\affiliation{Sackler School of Chemistry, Tel-Aviv University, Tel-Aviv, Israel}
\author{Mark G. Raizen}
\email{raizen@physics.utexas.edu}
\affiliation{Center for Nonlinear Dynamics and Department of Physics, The University of Texas at Austin, Austin, Texas 78712-1081, USA}

\begin{abstract}
We report the stopping of an atomic beam, using a series of pulsed electromagnetic coils. We use a supersonic beam of metastable neon created in a gas discharge as a monochromatic source of paramagnetic atoms. A series of coils is fired in a timed sequence to bring the atoms to near-rest, where they are detected on a micro-channel plate. Applications to fundamental problems in physics and chemistry are discussed.
\end{abstract}

\maketitle
In this Letter, we report the stopping of a supersonic beam using a scalable method applicable to any paramagnetic atom or molecule. We were originally inspired by a method of macroscopic velocity control, the coilgun, where a ferromagnetic projectile is accelerated and launched by passing through a sequence of electromagnetic coils which generate large pulsed magnetic fields with fast switching times. The University of Texas Center for Electromechanics, Sandia National Laboratories, and others have used a coilgun to accelerate and decelerate macroscopic samples, routinely achieving speeds of hundreds of meters per second \cite{hebner}. The idea is to use the coilgun method to decelerate free paramagnetic atoms in a beam, instead of firing a bulk projectile \cite{TheoMag}. The same approach has been independently pursued \cite{merkt1,merkt2}.\par
Shrinking the size of the projectile to the size of a single atom has several advantages. First, our solenoids can be miniaturized, allowing us to create high magnetic fields in a small volume with relatively low currents running through the solenoid windings. Furthermore, any atom that has a permanent magnetic moment can be slowed by the atomic coilgun. Most elements have unpaired electrons and a permanent magnetic moment in their ground state. In addition, our method is equally applicable to paramagnetic molecules.\par 
The paramagnetic atoms we slow are cooled in a supersonic expansion, which produces high flux, cold atomic and molecular beams by adiabatic expansion of high pressure gas through a small aperture into vacuum. The temperatures one can reach are in the range of several tens of millikelvin \cite{supersonics}. The enthalpy of the gas is converted into kinetic energy leading to high beam velocities, ranging from a few hundred to a few thousand meters per second, depending on the carrier gas mass and the source temperature. Atoms can be entrained into a supersonic beam by laser ablation or passing the beam through a vapor cell. Several methods have been used to slow supersonic beams; these include mechanical approaches such as spinning the source \cite{herschbach}, reflection off a receding wall \cite{rotor}, and crossed beam collisions \cite{billard}.  Atoms and molecules have been cooled via collisions with a cold background gas  \cite{buffergas}. Supersonic molecular beams have also been slowed via interactions with pulsed laser fields \cite{barker} as well as with pulsed electric fields \cite{stark,meijer-phase,stark_trap_meijer,stark_trap_ye}.\par
In a previous experiment, we demonstrated slowing of metastable neon from \zl{460}{m/s} to \zl{400}{m/s} in an 18 stage atomic coilgun \cite{1stMag}. Here we present results showing full control over the velocity of a supersonic beam of metastable neon with a 64 stage slower apparatus.\par
The principle of operation of the atomic coilgun is based on the Zeeman effect, in analogy to the pulsed electric field decelerator's use of the Stark effect \cite{stark,meijer-phase,stark_trap_meijer,stark_trap_ye}. In a magnetic field the atom's energy levels split into high-field and low-field seeking states. The low-field seekers gain potential energy as they move into the high magnetic field region at the center of an electromagnetic coil. When the atom reaches the top of the magnetic ``hill'' the magnetic field is suddenly switched off.  Due to conservation of energy, the kinetic energy an atom loses is equal to the Zeeman energy shift, 
\begin{equation}
\Delta E = g \mu_B m_J B
\end{equation}
where $g$ is the Land\'e factor, $\mu_B$ is the Bohr magneton, $m_J$ is the projection of the total angular momentum on the quantization axis and $B$ is the magnetic field. The same process can be repeated in another coil, gradually reducing the kinetic energy of an atom until it is brought to rest in the laboratory frame.\par
We generate the high magnetic fields needed for efficient beam deceleration with electromagnetic coils having 30 copper windings (\zl{0.5}{mm} wire diameter) and a bore of \zl{3}{mm}. Each solenoid is encased between two Permendur disks (\zl{10.2}{mm} diameter, \zl{3}{mm} thickness) and surrounded by a magnetic steel tube (inner diameter \zl{10.2}{mm}, outer diameter \zl{17.8}{mm}, length \zl{9.6}{mm}). This is a minor modification of our previous 18 stage prototype coil design and is similar to the solenoid used in our pulsed valve. An increase in the current passing through each of the coils from \zl{400}{A} in our prototype apparatus to the \zl{750}{A} currently used increases the magnetic field strength. We characterize the resulting magnetic field and the switching profile using the Faraday effect \cite{faraday}. Magneto-optical materials rotate the polarization of light by an amount proportional to the magnetic field parallel to the light. \par
We insert a terbium gallium garnet (TGG) crystal into the bore of our coil and monitor the rotation of a linearly polarized HeNe laser beam which we focus tightly (spot size \zlm{80}{m}) into the TGG crystal as we pulse the current in our coil. Using different length TGG crystals we calculate a peak field of \zl{5.2\pm 0.2}{T} in the center of the coil. The switching profile follows an initial exponential rise with a time constant of \zlm{19}{s}. When we switch off the current, the magnetic field falls linearly to \zl{20}{\%} of its peak value in \zlm{6}{s}. After this linear fall-off the field decays exponentially with a time constant of \zlm{17}{s} due to eddy currents induced in the Permendur disks by the rapidly changing field. The effect of eddy currents is partially reduced by a current counter-pulse.\par
An important consideration of the current driver design is its scalability since a large number of stages is needed to slow species with small magnetic moment to mass ratios. Although it is possible to switch current pulses in each channel with a high power insulated-gate bipolar transistor (IGBT), this solution is not practical due to the accumulating costs. In our setup we use only 8 IGBTs \cite{powerex} to independently control the discharge of the drive capacitor of each of the 64 channels. Each IGBT switches currents in 8 coils, while every coil in the set of 8 is isolated from one another with an inexpensive thyristor. We can easily expand our setup by adding more coils to every IGBT driver set.\par
\begin{figure}
\includegraphics[width=0.5\textwidth]{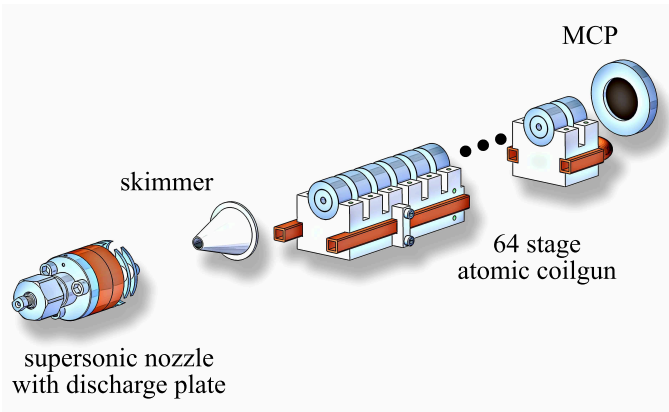}
\caption{\label{fig1} A descriptive drawing of our apparatus. Objects are to scale, but the distances between them are not.}
\end{figure}
We now describe our full apparatus as shown in figure \ref{fig1}.  The supersonic beam of  neon is created using an Even-Lavie supersonic nozzle \cite{uzi1, uzi2}. This valve is capable of pulses as short as \zlm{10}{s} FWHM, and produces intensities of \zl{10^{24}\:}{atoms/sr/s}. Since it is advantageous to start with a slow beam, we cool the nozzle to \zl{77}{K}. We use metastable neon because it has a magnetic moment, is simple to produce in a supersonic beam, and allows for easy and efficient detection. To produce metastable atoms, we apply a pulsed dc discharge between plates mounted \zl{5}{mm} from the exit of the nozzle. We find that a discharge pulse of \zlm{2.6}{s} produces a beam with a mean velocity of \zl{446.5\pm 2.5}{m/s} with a standard deviation of \zl{14.8\pm 0.2}{m/s}. This corresponds to a temperature of \zl{525\pm 10}{mK}.  The discharge produces metastable neon atoms in the 2p$^5$3s$^1$ electronic configuration, and we time our coil pulses to slow the $^3$P$_2$ $m_J=2$ state.  This state has a magnetic moment of $3\:\mu_B$ in both the low field Zeeman and high field Paschen-Back regimes.\par 
From the discharge, our beam travels to a \zl{5}{mm} diameter, \zl{50}{mm} long conical skimmer which is mounted \zl{300}{mm} from the exit of the nozzle. The center of the first coil is located \zl{250}{mm} from the skimmer base. The coilgun consists of 64 individually triggered solenoids, all mounted on  a monolithic aluminum support structure. This support serves both to align the coils and to provide a heat sink for the thermal energy generated in the coils by the pulsed current. We water-cool this aluminum support to keep our coils at a constant temperature. The solenoids are spaced \zl{14}{mm} apart (center to center) giving an overall length of \zl{900}{mm} for the slowing apparatus.  Following the coils, the beam then propagates to a micro-channel plate (MCP) which we use for detection \cite{ElMul}.  The MCP is mounted on a translation stage, allowing a direct means of measuring the speed of different components of the beam. The distance between the center of the last coil and the MCP can be varied between 40 mm and 90 mm. By mounting the MCP close to the exit of the slower, we lose very little flux due to angular divergence of the slow beam.\par 
By varying the pulse sequence timing, we can control the final velocity and flux of the resulting beam.  We use a field programmable gate array \cite{FPGA} to control the coil switching with a time resolution of \zl{100}{ns}.  To create a timing sequence for the coils we numerically simulate the trajectory of an atom traveling through our apparatus, and from this trajectory we calculate the switching times. The simulation uses magnetic fields whose spatial characteristics are calculated by finite element analysis, and a time profile obtained using the Faraday rotation measurement. One variable we adjust when generating the timing sequence is the position of an atom relative to a coil when we switch it off.  We refer to this position as a phase angle following the convention used in the Stark decelerator \cite{meijer-phase}, where $90^{\circ}$ refers to switching the coil off when the atom is centered in the coil being switched, and $0^{\circ}$ refers to switching when the atom is exactly between the coil to be switched and the previous coil. By changing the phase of a timing sequence, we are varying the amplitude of the magnetic field that an atom experiences when we switch the coil off.  If we switch at a lower phase, the atom experiences a lower field and is slowed less in each coil.  However, the advantage of going to a lower phase is that one gains a larger region of phase stability, which leads to a greater number of slowed atoms.\par 
\begin{figure}[t]
\includegraphics[width=0.5\textwidth]{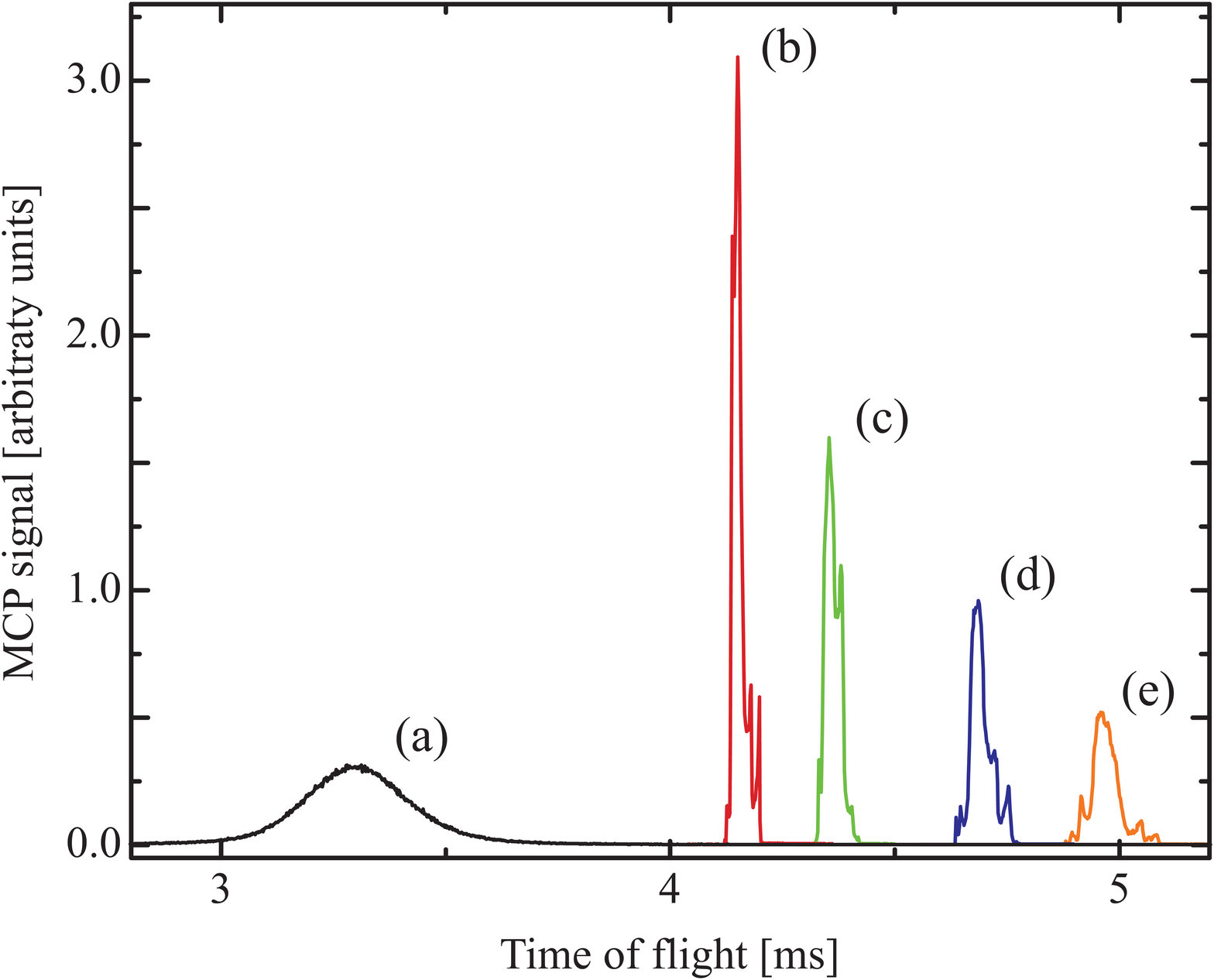}
\caption{\label{fig2}Time-of-flight measurements recorded with the MCP detector for different phases along with a reference beam (a). Each curve is an average over 20 shots at a repetition rate of \zl{0.075}{Hz}. The slowed curves show only the slowed portion of the beam for clarity. Beam velocities are (a) \zl{446.5}{m/s}, (b) \zl{222}{m/s}, (c) \zl{184.7}{m/s}, (d) \zl{142.7}{m/s}, and (e) \zl{109.9}{m/s}. Full results are summarized in table \ref{tab:1}.}
\end{figure}
\begin{figure}[t]
\includegraphics[width=0.5\textwidth]{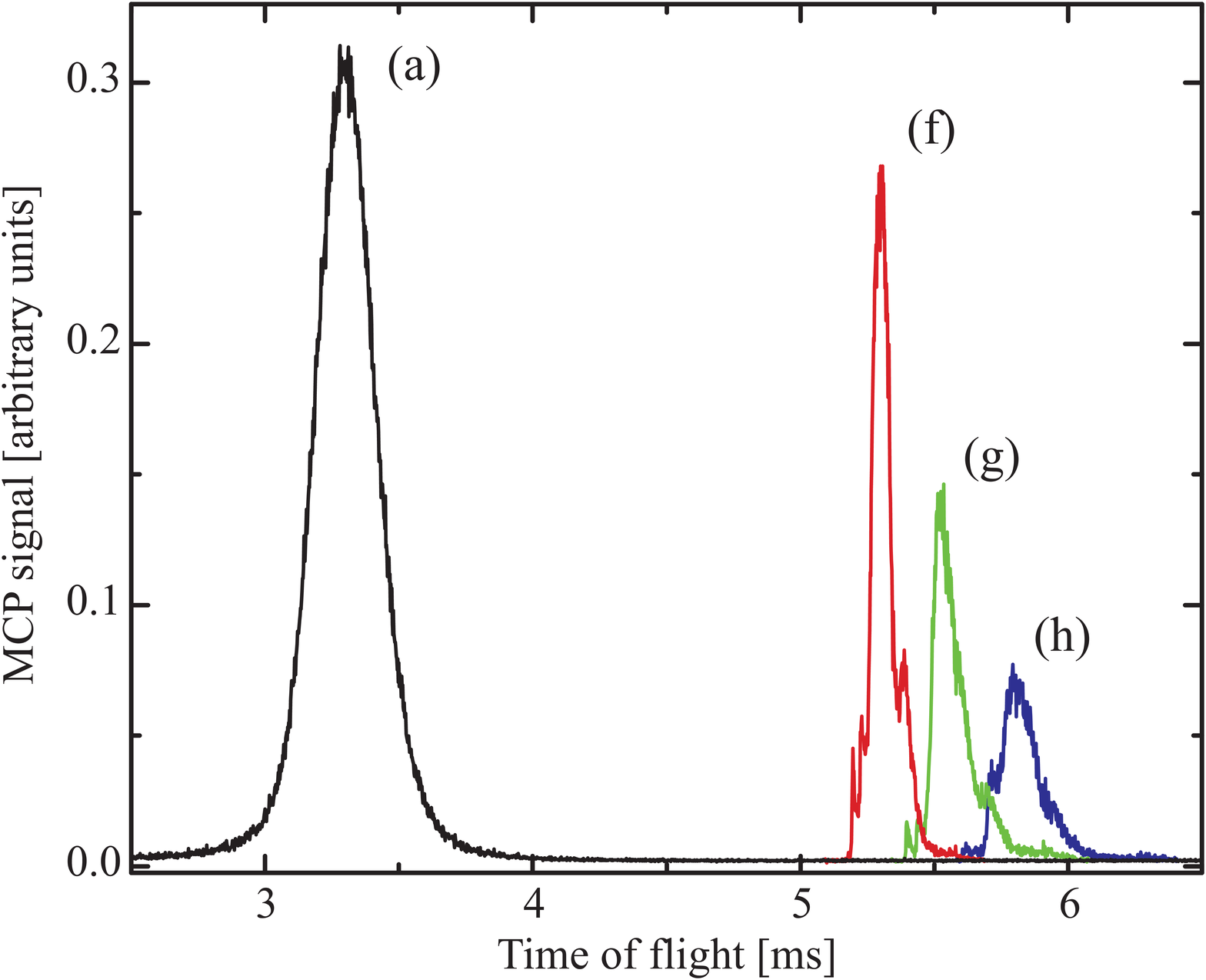}
\caption{\label{fig3}Time-of-flight measurements recorded with the MCP detector for 
different phases along with a reference beam (a). Each curve is an average over 20 shots at a repetition rate of \zl{0.075}{Hz}. The slowed curves show only the slowed portion of the beam for clarity. Beam velocities are (a) \zl{446.5}{m/s}, (f) \zl{84.1}{m/s}, (g) \zl{70.3}{m/s}, and (h) \zl{55.8}{m/s}. Full results are summarized in table \ref{tab:1}.}
\end{figure}

We now present results of slowing with phases between $36^{\circ}$ and $44^{\circ}$. We slow the atoms from \zl{446.5\pm 2.5}{m/s} to speeds as low as \zl{55.8\pm 4.7}{m/s}, which represents a removal of more than \zl{98}{\%} of the kinetic energy. Atoms can be stopped with a slightly higher phase and will be trapped at the exit of the slower in future experiments. In figures \ref{fig2} and \ref{fig3} we show slowing results for different phase angles, where the results are divided between high and low beam velocities. The time-of-flight profile of the slowed beam has structure that depends on the slowing phase. The anharmonicity of the magnetic potential causes the atoms to oscillate with different frequencies longitudinally in the slower, according to simulations. Thus the velocity distribution of the slowed peak will not be uniform, which is reflected in the features of the arrival time signal. Bethlem et al. have investigated this phenomenon in the Stark decelerator \cite{meijer-phase}.\par 
The slower has efficiencies ranging from \zl{2.0}{\%} to \zl{11.9}{\%} of the incident beam, depending on the final velocity. We determine the efficiency of our slowing apparatus by comparing the area under each slow peak to the geometrically scaled area under the reference beam, i.e. we normalize to the expected intensity of the reference beam at the entrance of the slowing apparatus. Note that the discharge does not selectively create atoms in the state we slow; the metastable population is distributed among the two long lived excited states, $^3$P$_2$ and $^3$P$_0$. We only slow the $m_J=2$ projection of the $^3$P$_2$ state, and the measured efficiencies do not take this into account. For example, if the projection states of $^3$P$_2$ neon are uniformly populated, the 2 \% slowing efficiency corresponds to 10 \% efficiency relative to the number of $m_J=2$ atoms.  These results are summarized in table \ref{tab:1}.\par 
\begin{table}[t]
\caption{\label{tab:1}Final velocities ($v_f$), temperatures ($T$) and efficiencies of the beams shown 
in figures \ref{fig2} and \ref{fig3}.}
\begin{ruledtabular}
\begin{tabular}{c|c|c|c}
 		&	 $v_f$ [m/s] 	&  $T$ [mK] 	&   efficiency [\%]	\\\hline
(a)	&	$446.5\pm 2.5$	&	$525\pm 10$	& 	 --					\\
(b)	&	$222  \pm 11 $	&	$108\pm 22$	&	$11.9\pm 0.5$	\\
(c)	&	$184.7\pm 7.6$	&	$184\pm 39$	&	$ 9.6\pm 0.4$	\\
(d)	&	$142.7\pm 9.1$	&	$117\pm 32$	&	$ 7.2\pm 0.3$	\\
(e)	&	$109.9\pm 5.4$	&	$147\pm 34$	&	$ 5.5\pm 0.2$	\\
(f)	&	$ 84.1\pm 3.1$	&	$ 79\pm 20$	&	$ 4.1\pm 0.2$	\\
(g)	&	$ 70.3\pm 7.4$	&	$ 92\pm 57$	&	$ 2.9\pm 0.1$	\\
(h)	&	$ 55.8\pm 4.7$	&	$106\pm 59$	&	$ 2.0\pm 0.1$	\\
\end{tabular}
\end{ruledtabular}
\end{table}
In summary, we demonstrate the operation of a scalable 64 stage atomic coilgun with peak magnetic field values reaching \zl{5.2}{T} in each deceleration coil. Full control over the velocity of a metastable neon supersonic beam is shown, starting from the initial \zl{447}{m/s} down to \zl{55.8}{m/s}, by tuning the phase angle. The atomic coilgun is a robust method, as it guides low-field seekers throughout the stopping process. The method can naturally bring the atoms or molecules to rest inside a magnetic trap located in a room-temperature chamber with excellent optical access. The atomic coilgun is ultimately simple, with small coils driven by discharging capacitors to produce the required magnetic fields.\par 
Further development of magnetic stopping, including the adiabatic deceleration of trapped atoms \cite{TheoMag} will be investigated.  We will examine methods of transferring the pre-slowed atoms to a static magnetic trap, with the goal of obtaining the high phase space densities provided by the supersonic beam itself.\par
We predict that the general nature of this method will open many new directions in physics and chemistry. One immediate goal in our group will be toward trapping of atomic hydrogen isotopes, H, D, and T. The prospects for precision measurement of beta decay and determination of the neutrino rest mass with trapped atomic tritium look very promising. This will be discussed in a future publication.\par
The application of our method to trapping of molecular radicals is very interesting. A magnetic slower could simultaneously trap several species which opens the door to a new regime of cold chemistry. For example, the fundamental combustion reaction $\text{H} + \text{O}_2\rightarrow\text{OH} + \text{O}$ can be studied in the quantum regime  \cite{combustion}.\par 
Possible technological applications include stopping magnetic atoms such as iron, nickel, and cobalt for controlled deposition of magnetic quantum dots and magnetic storage.  Another possible application would be in molecular beam epitaxy where the supersonic beam, together with magnetic control, could possibly reduce the use of toxic elements.\par 
 This work is supported by the  Army Research Office, the R.A.~Welch Foundation, the Sid W. Richardson Foundation  and the US National Science Foundation Physics Frontier Center.\par

\end{document}